\documentstyle[pre,aps]{revtex}

\begin{document}                             

\title{
Apparent Fractality Emerging from Models \\ 
of Random Distributions
}

\author{Daniel Hamburger$^{a}$\cite{dani},
Ofer Biham$^{a}$\cite{ofer} and David
Avnir$^{b}$\cite{david}}
\address{
$^{a}$
Racah Institute of Physics, The Hebrew University, Jerusalem 91904, Israel
}
\address{
$^{b}$
Institute of Chemistry, The Hebrew University, Jerusalem 91904, Israel
}

\maketitle

\begin{abstract}
\newline{}
The fractal properties of models of randomly placed $n$-dimensional spheres
($n$=1,2,3) are studied using standard techniques for calculating fractal
dimensions in empirical data (the box counting and Minkowski-sausage
techniques). Using analytical and numerical calculations it is shown that in
the regime of low volume fraction occupied by the spheres, apparent fractal
behavior is observed for a range of scales between physically relevant
cut-offs. The width of this range, typically spanning between one and two
orders of magnitude, is in very good agreement with the typical range observed
in experimental measurements of fractals. The dimensions are not universal
and depend on density. These observations are applicable to spatial, temporal
and spectral random structures. Polydispersivity in sphere radii and
impenetrability of the spheres (resulting in short range correlations) are
also introduced and are found to have little effect on the scaling properties.
We thus propose that apparent fractal behavior observed experimentally over a
limited range may often have its origin in underlying randomness.
\end{abstract}

\pacs{64.60.Ak,05.40.+j}

\section{Introduction}
In recent years the study of fractal structures has been an active field of
research both theoretically and experimentally
\cite{Mandelbrot,Feder,Bunde,Stanley,Avnir:book,Takayasu,Pynn,deGennes,Bara}.
In theory, a variety of algorithms and dynamical models which produce fractal
sets have been introduced. Typically, in these models one can define an
asymptotic limit in which the set exhibits fractal behavior on an arbitrarily
broad range of length scales.  One can then approach this limit by a process
of gradual refinements of the set, which may involve either an increase in
system size or decrease in the minimal object size.  In the case of {\it
empirical fractals} observed experimentally the situation is different.  For
these fractals the range over which they obey a scaling law is restricted by
inherent upper and lower cutoffs.  In most experimental situations this range
may be quite small, namely not more than one or two orders of magnitude
\cite{me:D}.  Nevertheless, even in these cases the fractal analysis condenses
data into useful relations between different quantities and often provides
useful insight \cite{Feder,Bunde,Stanley,Avnir:book}.

Motivated by the yet largely inexplicable abundance of experimentally observed
fractals, we consider in this paper the apparent fractal properties of systems
which are governed by uniformly random distributions.  The choice of random
systems is justified by the abundance of randomness in nature, and by the fact
that uniform randomness is a convenient limit, on top of which correlations
can be introduced as perturbations.  Although a purely random system cannot be
fully scale invariant, it may, as we show below, display apparent fractality
over a limited range.  The width and the cut-off values of this range seems to
be in good agreement with the typical width and typical cut-offs observed in
experimental measurements (between one and two decades), unlike the case with
models which are inherently scale free.
 
To illustrate these ideas we consider a model in which $n$-dimensional
($n=1,2,3$) spheres
of diameter $d$ are randomly distributed in an $n$-dimensional space in the
regime of low volume fraction occupied by the spheres.  In three dimensions
(3D) our definitions coincide with ordinary spheres, while for $n=2$ (2D) we
consider disks, and for $n=1$ (1D) rods of length $d$.  In the basic model the
positions of the spheres are uncorrelated and they are thus allowed to
overlap. We then extend the model to the case where there is a distribution of
sphere radii and examine the effect of this distribution on the fractal
properties. We also examine a version of the model in which spheres are not
allowed to overlap, thus introducing short range correlations between sphere
positions.  This class of models may approximately describe spatial
distribution of objects such as craters on the moon, droplets in a cloud and
adsorbates on a substrate. In particular, the one dimensional model may
describe the level distribution in energy spectra of quantum systems and the
zeroset of random temporal signals.  Therefore, such models may be at the root
of empirical observations of fractals in experiments and data analyses,
dealing with processes governed mainly by randomness \cite{me:D}.  As
fractality is usually revealed by applying various resolution analyses, we
also address the question of whether our observations are method-dependent.

Two of the most commonly employed resolution analysis methods are the
box-counting (BC) and Minkowski sausage (MS) techniques
\cite{Mandelbrot,Falconer}.  The fractal properties of these models are
studied here both analytically and numerically, within the BC and MS
frameworks.  The analytical solution is exact except for the case of
impenetrable spheres in $n>1$ dimensions where some approximations were
needed.

In the box-counting (BC) procedure one covers the embedding space by
non-overlapping boxes of linear size $r$, and then counts the number of boxes
$N(r)$, which have a non-empty intersection with the (fractal) set.  A fractal
dimension $D_{BC}$, is declared to prevail at a certain range, if a relation
of the type

\begin{equation}
N_{BC}(r) \sim r^{-D_{BC}}
\label{eq:N=rtoD}
\end{equation}

\noindent holds, or equivalently, if the slope of the log-log plot

\begin{equation}
D_{BC} = - {\rm slope}\:\{\log r, \log[N_{BC}(r)] \}
\label{eq:D_BC}
\end{equation}

\noindent is found to be constant over that range. In the MS case
\cite{Falconer} one draws an $n$-dimensional ($n$=1,2,3) Minkowski-sphere
(M-sphere) of radius $R$ around each point in the set under consideration, and
calculates how the volume $V$ of the union of all spheres changes under a
variation of $R$.  The set is considered fractal with a dimension $D_{MS}$,
over a range of scales, if

\begin{equation}
V(R) \sim R^{n-D_{MS}} ,
\label{eq:V=RtoD}
\end{equation}

\noindent or equivalently, if

\begin{equation}
D_{MS} 
= - {\rm slope}\:\{\log R, \log[V(R)/R^n] \} 
\label{eq:D_M}
\end{equation}

\noindent is constant within this range.  The BC and MS methods are known to be
identical from the mathematical point of view \cite{Falconer}, where the
limits $r,R \rightarrow 0$ can be taken. Their equivalence from the physical
point of view, under the constraints of finite cut-offs, is not obvious.
While it is clear from Eqs.(\ref{eq:N=rtoD})-(\ref{eq:D_M}) that both methods
employ resolution analysis, in which the number of occupied ``resolution
units'' (boxes or spheres) has to be determined as a function of the
resolution magnitude (box-length or M-sphere-radius), there are slight
differences between the two methods, which are due to both geometrical
differences and the presence of cut-offs.

Using a resolution analysis one should be aware of the existence of physical
cut-offs.  This is especially important in the structures considered below
which are not fractal in the rigorous mathematical sense.  We will show that
the log-log plots of the functions $N(r)$ and $V(R)$ display linear behavior
between these physical cut-offs. The slope in this range can be interpreted as
an FD.  The existence of this scaling behavior between physically relevant
cutoffs is a central motivation for the study of the random models presented
below.

The paper is organized as follows: in Section \ref{basic-model} we consider
the basic model of penetrable spheres where all the spheres are of equal size,
and calculate the BC function. From the analysis of this function we obtain
the nontrivial linear range and extract the apparent FD. The generality of
this model is then demonstrated using an information theory argument. In
Section \ref{poly} we generalize the analysis to the case where the spheres
are not equally sized but exhibit various size distributions and examine the
effect on the apparent fractal behavior.  Certain correlations are then
introduced in Section \ref{impenetrable} where the case of impenetrable
spheres in considered and the conclusions are given in Section
\ref{conclusions}.

\section{The Model of Randomly Distributed Spheres}
\label{basic-model}

\subsection{The 1D Model}
\label{model}

In this model \cite{me:D} $M$ rods of length $d \ll 1$ are randomly placed on
the unit interval (Fig.\ref{fig:rods}) \cite{random-comment1} such that the
positions of the rod centers are chosen from a uniform random distribution.
The rods are mutually penetrable, namely overlaps are allowed.  The BC
function $N(r)$, will now be derived.  This function gives the number of
boxes, for given box size $r$, which have a non-empty intersection with the
set.  For a large enough number of rods (of the order of 100 in the present
case), the deviations from the expectation value $\langle N(r) \rangle$ are
negligible and edge effects are unimportant.  Let $p$ denote the probability
that a box of size $r$ intersects a rod of length $d$.  Then for a total of
$r^{-1}$ boxes,

\begin{equation}
\langle N(r)\rangle = {p \over r}.
\label{eq:N}
\end{equation}

\noindent Following Refs.\cite{Weissberg,Torquato:3} define the probability
$q_1$ that after random placement of the first rod, a given box remains
unintersected.  Neglecting edge effects, this implies that the center of the
rod must be at least a distance of $d/2$ away from either edge of the box
(Fig.\ref{fig:rods}(c)).  A total length of $r+2(d/2)$ is, therefore,
unavailable for placement without intersection.  For a uniformly random
distribution it follows that:

\begin{equation}
q_1 = 1-(r+d) .
\label{eq:q1}
\end{equation}

\noindent The next rods are placed independently, which means that after
placement of $M$ rods, the probability that the box is still unintersected is

\begin{equation}
q = (1-(r+d))^{M} .
\label{eq:qNd}
\end{equation}

\noindent Finally, the probability of at least one intersection is $p = 1-q$,
and therefore:

\begin{equation}
\langle N(r)\rangle = {1\over r} \left[ 1-\left(1-(r+d) \right)^{M} \right] .
\label{eq:<N>}
\end{equation}

\noindent This is the one dimensional BC function for randomly adsorbed,
mutually penetrable rods \cite{me:D}.  Numerical simulations of the model
along with the theoretical prediction of Eq. (\ref{eq:<N>}), are shown in
Fig.\ref{fig:Nresults}. The excellent agreement is evident. The parameters
$d$ and $M$ are independent, and are limited only by the restriction $\eta
\leq 1$, where $\eta \equiv M d$ is the coverage of the line. We find that
for large M (here $M>100$) changing $d$ and $M$ while keeping $\eta$ fixed,
merely translates rigidly the BC function in the log-log plane.

In Fig.\ref{fig:Nresults.reg}, the BC function is shown in a range between
cut-offs, together with a linear regression. This is the range which is
commonly measured in experimental work. The special significance of these
cut-offs is discussed below. Notice the nearly linear appearance,
extending over close to 1.5 decades.

\subsection{Extension to 2D and 3D}
\label{2D3D-extension}

In the two dimensional model one places {\em disks} of diameter $d$ on the
unit square, so that for a given box of area $r^2$ to remain unintersected, no
disk center may fall within the area shown in Fig.\ref{fig:rods}(d). Thus, a
total area of $r^2 + 4({1 \over 4} \pi \,(d/2)^2) + 4(r\,d/2)$ is excluded for
placement of the first disk center.  Therefore in this case the probability of
the box to remain empty is:

\begin{equation}
q_1 = 1-\left(r^2 + {1\over 4} \pi \,d^2 + 2r\,d \right).
\label{eq:q1-2D}
\end{equation}

\noindent The next disks are placed independently, leading to:

\begin{equation}
\langle N(r)\rangle = {1 \over r^2} \left[ 1- \left( 1- \left( r^2 + 2r\,d +
{1\over 4} \pi \,d^2 \right) \right)^{M} \right].
\label{eq:<N>-2D}
\end{equation}

\noindent Similarly, for 3D one considers independent placement of spheres,
and the excluded volume for sphere centers is that formed by convolution of a
cube of side $r$ and a sphere of diameter $d$, which is $r^3+6(r^2 \,
d/2) + 12(r\,{1 \over 4} \pi (d/2)^2) + 8({1 \over 8}\, {4 \over 3} \pi\,
(d/2)^3)$. One obtains then

\begin{equation}
q_1 = 1-\left(r^3 + 3r^2 \,d + {3\pi \over 4} r \,d^2 + {\pi \over 6}d^3
\right),
\label{eq:q1-3D}
\end{equation}

\noindent and

\begin{equation}
\langle N(r)\rangle = {1 \over r^3} \left[ 1- \left( 1-\left(r^3 + 3r^2 \,d +
{3\pi \over 4} r \,d^2 + {\pi \over 6}d^3 \right) \right)^{M} \right].
\label{eq:<N>-3D}
\end{equation}

\subsection{Cut-offs, Range of Linearity, and the Fractal Dimension}
\label{penetrable-FD}

\subsubsection{Cut-offs}
We will now study the BC function $N(r)$ and examine the possibility of
fractal-like behavior.  For simplicity we will first concentrate on the 1D
case, but the conclusions apply to 2D and 3D as well.

For mathematical fractals displaying full scale-invariance, the log-log plot
of $N(r)$ vs. $r$ can form a straight line with a fractal slope over an
unlimited range of scales in the asymptotic limit.  For the set we consider,
this is clearly not the case as in Fig.\ref{fig:Nresults}: there are ``knees''
beyond which the slope approaches 1.  This is due to the existence of lower
and upper cut-offs, $r_0$ and $r_1$ respectively.  These cut-offs correspond
to relevant physical limits of observation.  Here, the smallest feature is of
size $d$, and so the finest resolution is of that size.  No additional
information is obtained by reducing $r$ below $r_0=d$ , where $D$ approaches
the trivial limit of $1$ as $r \rightarrow 0$. To see why this is so, suppose,
for convenience, that the center of each rod is located at a point connecting
two adjacent boxes: then halving the box doubles the number of intersected
boxes if $r<d$, with the result that $D$ must approach the limit of $1$ as $r
\rightarrow 0$. When the centers of the rods are located arbitrarily, the
lower cut-off will not be sharply located at $d$. Nevertheless,

\begin{equation}
r_0=d
\label{eq:r_0}
\end{equation}

\noindent is a good estimate for it. As for the upper cut-off $r_1$, it is the
size beyond which practically all boxes intersect at least one rod, where
again $D\rightarrow 1$. This happens when the boxes are larger than the
average gap between rods:

\begin{equation}
r_1=1/M-d
\label{eq:r_1},
\end{equation}

\noindent which is therefore an approximate upper cut-off.  We thus have the
approximate range $r_0 < r < r_1$, where the measurement is properly tuned to
measure inherent scaling behavior, if it exists.  For scaling behavior to be
observed, there must be a minimal range of apparent linearity
(Eq.(\ref{eq:D_BC})). The size of such a range and the extent of linearity
displayed by the BC function in it are considered below.

\subsubsection{Range of Linearity}
\label{range}
The standard experimental procedure is to apply a linear regression analysis
on the log-log presentation of the scaling range.  The linear regression line
is constructed to go through the inflexion point $(r_i, N(r_i))$ of the
log-log plot of Eq.(\ref{eq:<N>}).  The dependence of the range of linearity
$\Delta$ on the coefficient of determination, ${\cal R}^2$, (measuring the
quality of the linear description) is then explored.

The range of linearity is approximately given by
$\Delta=2(\log(r_i)-\log(r_0))$.  Applying a linear regression analysis
on the log-log plot of the BC function, we evaluated the slopes and actual
ranges of linearity under these constraints, with different values of ${\cal
R}^2$ imposed. Typical results are shown in Fig.\ref{fig:range}.  For
instance, about two decades of linear behavior can be observed for a required
value of ${\cal R}^2$ of below 0.97. Examples of experimentally observed
fractal objects exhibiting several orders of magnitude of linearity are rare;
the vast majority of reported fractality spans 1-2 orders of magnitude
\cite{Feder,Bunde,Stanley,Avnir:book,Takayasu,Pynn,deGennes,me:D,D:histogram}.
It is important to emphasize that, in fact, we are mimicing in our
calculations the common practice of searching and reporting FD's in empirical
data.

The convention for the smallest meaningful scaling range is {\em one
decade}\cite{Pfeifer-Avnir:book}. Given this, a simple argument yields the
maximum allowed coverage: By using the estimates above for the cut-offs,
$r_0=d$ and $r_1 = 1/M-d = d(1-\eta)/\eta$, one observes that a range $\Delta
= \log(r_1)-\log(r_0) = \log[(1-\eta)/\eta]$ of at least one decade, requires
that: $\eta<1/11$. It follows that the apparent fractality in our model is
{\em restricted to $\eta \leq 0.1$}. The range between the
cut-offs grows as the coverage is {\em decreased},
as observed 
in Fig.\ref{fig:Nresults}.

In addition to the width of the range between the cutoffs, the quality
of the linear fit within this range,
measured by ${\cal R}^2$
should also be considered.  
One can limit the range of linearity by
imposing a lower bound on  ${\cal R}^2$:
obviously, the range decreases as ${\cal R}^2$
increases (Fig.\ref{fig:range}). 
Also note from this Figure, that for
a given range, the quality of linearity 
grows as the coverage is {\em increased}, or as the
slope of the BC function between the 
cut-offs (i.e., the FD) approaches 1
(Fig.\ref{fig:Nresults}). 
This happens because of the smooth merging with the
slope beyond the cut-offs, which is trivially 1. 
We thus conclude that the two cut-offs limit the width of the linear
range for high coverage while the ${\cal R}^2$ criterion limits it 
for low coverage. As a result, the range of scales in which we observe
apparent fractality is typically between one and two orders of magnitude.

\subsubsection{Fractal Dimension}
The apparent FD shown in Fig.\ref{fig:D}, for an imposed values of ${\cal
R}^2=0.995$, rises monotonically from 0 as more rods are added.  This is an
important aspect of the model: it does not predict a universal (specific) FD,
but the whole allowable range of FD values. The regression results are further
compared in Fig.\ref{fig:D} to an analytical equation, obtained by calculating
the logarithmic derivative of $N(r)$ at the {\em estimated} middle point $r_e
= \sqrt{r_0 r_1}$:

\begin{equation}
D_{BC}^{(1)} = {d(\log[N(r)]) \over d(\log[1/r])}\left|_{r=r_e}\right. =
1-\sqrt{\eta(1-\eta)} {{\left( 1-{{\eta+\sqrt{\eta(1-\eta)}} \over M}
\right)^{M-1}} \over {1-{\left(1-{{\eta+\sqrt{\eta(1-\eta)}} \over M}
\right)^{M}}}} .
\label{eq:Dresult}
\end{equation}

\noindent We use an estimate for the middle point ($r_e$) rather than the exact
result ($r_i$), since $r_i$ cannot be given analytically. The almost symmetric
S-shape of the log-log plot in the scaling region assures that $r_e$ is a good
estimate for $r_i$.  As seen in Fig.\ref{fig:D}, the FD predicted by
Eq.(\ref{eq:Dresult}) is an accurate lower bound to the regression result.

By using $(1+x/N)^N \rightarrow e^x$ as $N \rightarrow \infty$,
Eq.(\ref{eq:Dresult}) may be simplified in the ``thermodynamic limit'', while
keeping $\eta$ finite. One then obtains:

\begin{equation}
D_{BC}^{(1)} = 1-{\sqrt{\eta(1-\eta)} \over {\exp \left( \eta +
\sqrt{\eta(1-\eta)} \right)-1}} .
\label{eq:Deta}
\end{equation}

\noindent Notably, the expression for $D_{BC}^{(1)}$ depends on $\eta$
alone. In the limit of small $\eta$, one can further simplify
Eq.(\ref{eq:Deta}) and obtain

\begin{equation}
D_{BC}^{(1)} \approx \left(\eta \over {1-\eta}\right)^{1/2}, \ \ \ \ \ \ \eta
\ll 1.
\end{equation}

\noindent Expressions for $D_{BC}^{(2)}$ and $D_{BC}^{(3)}$ may be derived from
Eqs.(\ref{eq:<N>-2D}),(\ref{eq:<N>-3D}) for the 2D and 3D cases. These, as
well as a discussion of their cut-offs and range of linearity are deferred to
Sec.\ref{poly-BC}, where a more general model is treated.

\subsection{Absence of Correlations}
\label{correlations}
Since fractal objects typically exhibit some correlations, one might wonder
whether the finding of an apparent FD in our model, is also due to some hidden
correlations within a certain window of resolution.  For instance, it might
seem plausible that the finite extent of the rods introduces a correlation,
for if a point on the line belongs to a certain rod, then a point at a
distance $x<d$ is likely to belong to the same rod. However, correlations at a
scale $x<d$ below the lower cut-off are, rightly so, typically not
measured. Furthermore, as we show next, the correlation exponent for $x>d$
vanishes. This will prove that the emergence of an FD in the penetrable rods
case is {\em not} due to the presence of correlations, but entirely due to the
scaling displayed by the underlying uniform distribution.

A correlation exponent, $\nu$, is associated with a given set if a correlation
function $c(x)$ has the following power-law form over a sufficiently large
range:

\begin{equation}
c(x) \sim x^{-\nu} .
\label{eq:corr}
\end{equation}

\noindent An exact expression for this function may be derived for penetrable
rods. Let $\theta(x_0)$ be the local density at the point $x_0$ on the
line. That is, $\theta(x_0) = 1$ if $x_0$ belongs to a rod, and zero
otherwise. The correlation function is defined as:

\begin{equation}
c(x) = \langle \theta(x_0) \, \theta(x_0+x) \rangle
\label{eq:c}
\end{equation}

\noindent where $\langle \cdots \rangle$ denotes either an ensemble average (at
fixed $x_0$), or an average over $x_0$. Assuming ergodicity, we choose to
calculate the latter. Only pairs of points $(x_0,x_0+x)$ such that both
$\theta(x_0)$ and $\theta(x_0+x)$ are $1$ contribute to the average. We thus
require the simultaneous probability:

\begin{equation}
P(x,x_0) \equiv {\rm Pr}\left[ (\theta(x_0+x)=1) \,\cap \, (\theta(x_0)=1)
\right] .
\label{eq:prob}
\end{equation}

\noindent The correlation function is then:

\begin{equation}
c(x) = \int_0^1 dx_0 \: \theta(x_0) \, \theta(x_0+x) \, P(x,x_0) ,
\label{eq:c-integral}
\end{equation}

\noindent where one can set $\theta(x_0) = \theta(x_0+x) = 1$. Neglecting edge
effects, $P(x,x_0)$ does not depend on $x_0$, since we are considering a
statistically homogeneous medium (or stationary process).  In what follows,
therefore, $x_0$ is considered to be any convenient reference point, though
none of the results depends on its location. With this general choice of
$x_0$, one obtains from Eq.(\ref{eq:c-integral}):

\begin{equation}
c(x) = P_{x_0}(x) = {\rm Pr}\left[ (\theta(x_0+x)=1) \,\cap \, (\theta(x_0)=1)
\right] .
\label{eq:c-P}
\end{equation}

\noindent To evaluate $P_{x_0}(x)$, consider a modification of the argument
that led to Eq.(\ref{eq:qNd}) : pick two ``test-points'' at random on a line
of length $L$ and denote their respective positions $x_0$ and $x_0+x$. Now
consider randomly placing rods of length $d$ on the line. The events which are
complementary to {\em both} points $(x_0,x_0+x)$ being occupied, are (1) that
at least one is unoccupied (with probability $Q_1$), and (2) that both are
unoccupied ($Q_2(x)$). Clearly $Q_1$ already accounts for $Q_2(x)$, so that:

\begin{equation}
P_{x_0}(x) = 1 - [2 Q_1 - Q_2(x)] .
\label{eq:P-q}
\end{equation}

\noindent The evaluation of $Q_1$ is a repetition of the argument leading to
Eq.(\ref{eq:qNd}), with a vanishing box size ($r=0$). Thus:

\begin{equation}
Q_1 = \left(1-d\right)^{M} \rightarrow  e^{-\eta}.
\label{eq:Q1}
\end{equation}

\noindent when $M \rightarrow \infty$. $Q_2(x)$ requires that no rod center is
placed within a distance smaller than $d/2$ to either test-point (see
Fig.\ref{fig:rods}(e,f)). When $x>d$, this clearly excludes a total length of
$2d$ from possible placement of rod centers. When $x<d$, a length of only
$x+d$ is excluded, due to overlap of the two inner segments. In all other
respects the argument leading to Eq.(\ref{eq:qNd}) (with $r=0$) is again
repeated, so that:

\begin{equation}
Q_2(x) = \left(1-(d + s(x))\right)^{M} \rightarrow e^{-\rho (d+s(x))} ,
\label{eq:Q2}
\end{equation}

\noindent 
where $\rho^{-1} = 1/M$ is the average distance between rod centers, and

\begin{eqnarray}
s(x) = \left\{ \begin{array}{ll}
	d          & \mbox{: $x \geq d$} \\
	x          & \mbox{: $x < d$ .}
	\end{array}
\right.
\label{eq:s(x)}
\end{eqnarray}

\noindent Combining Eqs.(\ref{eq:c-P}-\ref{eq:s(x)}), one finally obtains for
the correlation function:

\begin{eqnarray}
c(x) = \left\{ \begin{array}{ll}
	1-2 e^{-\eta}+ e^{-2\eta}               & \mbox{: $x \geq d$} \\
	1-2 e^{-\eta}+ e^{-\eta} e^{-\rho x}    & \mbox{: $x < d$ .}
	\end{array}
\right.
\label{eq:c-final}
\end{eqnarray}

\noindent Thus for $x \geq d$, $c(x)$ is constant, i.e., {\em the correlation
exponent vanishes} (a result obtained numerically by Meakin and
Deutch\cite{Meakin:3}).  However, the non-trivial range of the similarity
dimension {\em starts} at $d$, and hence cannot possibly be due to
correlations. These exist, not surprisingly, for $x<d$, but even there they
decay {\em exponentially} (with a characteristic length of $1/M$), and not as
a power-law, as required for fractality measured by the scaling of correlation
functions.

This completes the demonstration that the non-trivial result we obtained for
the similarity dimension of penetrable rods is {\em not} due to
correlations. The elements of this demonstration are applicable also to the
more general model presented in the next section and are not repeated there:
the absence of correlations holds for all penetrable cases treated in this
study.

\subsection{Generality of the Model}
\label{generic}

In this section arguments are presented showing that the model of randomly
placed, mutually penetrable spheres introduced in Sec.\ref{basic-model}, is
very general in the sense that it is a prototype of a much larger family
random processes.  Being minimalistic in assumptions, the model is in fact a
{\em generic} one for random processes.  An information-theoretical approach
is now employed to show this, using the fact that the only assumption entering
the model is the {\em knowledge of a mean quantity}. An important feature of
the information-theoretical approach is that it extends the model from spatial
to {\em temporal} random sequences, and even to energy-level distributions
\cite{Pfeifer:energy-levels}. Thus, in an adsorption process the relevant mean
quantity is the average density of adsorbates and one seeks the distribution
of nearest-neighbor distances. An equivalent situation is that in which one
knows the average period of a time signal and is interested in the
distribution of intervals between successive crossings of the time axis (the
``zeroset''). A finite width $d$ is then the signal width at the
crossings. Yet another case is that of spectral-level distribution (where $d$
is the uncertainty band width), for which it is well known that spacings of
quantum energy levels in classically non-chaotic systems are exponentially
distributed \cite{berry}. Indeed, it will be shown that an exponential
distribution of the intervals between the centers of the spheres is an
inherent characteristic of our model. It shall then become clear that spatial,
temporal and spectral random processes fall into the same class. The spatial
and temporal cases differ only in that a time-process must be {\em ordered} on
the time axis, while the positions of ``events'' on a spatial axis are not
necessarily ordered. There are, however, $N!$ possible arrangements in the
spatial case, obtained by interchanging the labels of events on the spatial
axis. The time-process may be considered as the one ordered set out of all
these permutations. Relabeling is all that is required to map a spatial
process to one in time, and clearly this does not affect the statistics of
positions or intervals. Furthermore, it is well-known \cite{Mandelbrot} that
the FD of the actual time signal can be deduced from that of its
zeroset. Relying on the equivalence just discussed, we choose in what follows
for concreteness to work on the spatial process of placement, but one should
keep in mind that the discussion applies just as well to time-processes and
energy-level spacings.

\subsubsection{Information Theory Argument}
Consider now the derivation of the distribution satisfying the assumption of
knowledge of the mean. In the present case this is the known average placement
density

\begin{equation}
\langle x \rangle = {M \over L}
\label{eq:x-ave}
\end{equation}

\noindent (or equivalently, this may represent the average period of a time
signal.) The arguments presented here for 1D are again easily generalized to
2D and 3D. Following a standard information theory argument \cite{Reza}, the
resulting distribution $P(x)$ of distances $x$ between neighboring adsorbate
centers is obtained by maximizing the missing information, $S$,

\begin{equation}
S = - \int_0^{\infty} P(x) \, \ln P(x) \,dx ,
\label{eq:S}
\end{equation}

\noindent where the constraint of knowing $\langle x \rangle$ can be written
as:

\begin{equation}
\langle x \rangle = \int_0^{\infty} x\, P(x) \,dx .
\label{eq:<x-ave>}
\end{equation}

\noindent To this one must add the normalization constraint

\begin{equation}
\int_0^{\infty} dx\: P(x) = 1 .
\label{eq:normalization}
\end{equation}

\noindent Using Lagrange multipliers $\lambda$ and $\mu$, the maximization of
$S$ can then be written in terms of a functional $F$ as:

\begin{equation}
F[P(x)] = - \int_0^{\infty} dx\: \{ P(x) \, \ln P(x) + \lambda \,P(x) + \mu
\,x\, P(x) \} ,
\label{eq:F}
\end{equation}

\noindent whose variational derivative is:

\begin{equation}
\delta F = - \int_0^{\infty} dx\: \{\ln P + 1 + \lambda + \mu \,x \} \delta P .
\label{eq:dF}
\end{equation}

\noindent The arbitrariness of $\delta P$ then requires the vanishing of the
term in curly brackets, or:

\begin{equation}
P(x) = e^{-(1+\lambda)} e^{-\mu x} .
\label{eq:P}
\end{equation}

\noindent Inserting this into the constraint equations,
Eqs.(\ref{eq:<x-ave>}),(\ref{eq:normalization}) yields the distribution of
nearest-neighbor distances

\begin{equation}
P(x) = {1 \over \langle x \rangle}\, e^{-{x \over \langle x \rangle}} .
\label{eq:P-final}
\end{equation}

\noindent The appearance of an exponential distribution is not surprising: it
is the Maxwell distribution when the average energy is given as the constraint
in the canonical ensemble; or it shows up as the distribution of
time-intervals between successive radioactive decays, where the mean life-time
acts as the known constraint. It remains to be shown, returning to the
adsorption-language, that given this exponential distribution of intervals
between the centers of the rods, the adsorbate {\em positions} are {\em
uniformly} distributed.

\subsubsection{The Position Distribution}
It is a standard exercise in probability theory to show that a uniform
distribution of positions leads to an exponential distribution of intervals
(see e.g. Ref.\cite{Brownlee}, or recall the argument for the time of flight
of a particle undergoing random collisions with a given mean free path).  It
is now shown that the opposite holds as well, namely that the exponential
distribution of intervals, derived above from an information-theory argument,
leads to a uniform distribution of positions.  The argument that led to the
derivation of the exponential distribution of intervals,
Eq.(\ref{eq:P-final}), assumed that only the mean distance between points is
known, and that this is the only parameter of relevance. Therefore, it was in
fact implicitly assumed that successive placements are independent (for
otherwise additional constraints should have appeared, reflecting the
dependence of the distribution of intervals on the number of previous
placements). Let $P(x)$ denote the probability density of finding a point
between $x$ and $x+dx$ after a single placement. Given that there is a point
at $x_0$, consider the conditional probability density $g(x_0+x | x_0)$ of
finding the nearest neighboring point between $x_0 + x, x_0 + x+dx$. This can
be expressed as:

\begin{equation}
g(x_0+x | x_0) = {1 \over \langle x \rangle}\, e^{-{x \over \langle x
\rangle}} P(x_0 + x) dx ,
\label{eq:gx0xx0}
\end{equation}

\noindent where $\exp(-x/\langle x \rangle)/\langle x \rangle$ is the
probability density of finding a gap of length $x$. But since this exponential
probability depends only on the (non-negative) {\em distance} between
neighboring points, it is clear that nothing prevents interchanging the roles
of $x_0$ and $x_0 + x$, i.e., it must hold that:

\begin{equation}
g(x_0+x | x_0) = g(x_0 | x_0+x) ,
\label{eq:g=g}
\end{equation}

\noindent or, explicitly:

\begin{equation}
{1 \over \langle x \rangle}\, e^{-{x \over \langle x \rangle}} P(x_0 + x) dx = 
{1 \over \langle x \rangle}\, e^{-{x \over \langle x \rangle}} P(x_0) dx_0 .
\label{eq:g=g-explicit}
\end{equation}

\noindent One is at liberty to choose $dx=dx_0$, so that $P(x_0)=P(x_0+x)$,
which holds for every $x$. Therefore $P(x)$ is {\em constant}, i.e., the
positions are uniformly distributed. This result followed from the exponential
distribution which was derived under the minimalistic assumption of knowledge
of the mean of a relevant property. Drawing on the generality of this
derivation, we conclude that {\em {a uniform random distribution of adsorbate
centers is a generic model of random processes in space and time}}, where one
only assumes knowledge of the mean. A uniform distribution of adsorbate
centers is, however, exactly what was assumed in the adsorption model in
Sec.\ref{basic-model}, for which apparent fractality was detected. Fractality
may therefore be expected for any other random system which can be
characterized by its mean.

One may further employ the above information-theory formalism in order to
derive the distributions appropriate to knowledge of higher moments, if
correlations are present in the system.

\section{Polydispersed Mutually Penetrable Spheres}
\label{poly}

\subsection{The Model}
The basic model introduced in Sec.\ref{basic-model} is now generalized, by
allowing polydispersivity in radii. That is, we consider a model of randomly
placed $n$-dimensional spheres with a {\em distribution of radii} $P(a)$. The
radii are assumed to be chosen independently from $P(a)$. In 1D, this may,
e.g., represent a random spectrum with levels exhibiting a distribution of
lifetimes. In 2D, such systems may approximately describe, for example, the
formation of metal clusters on metal surfaces
\cite{Kern:1,Kern:2,Comsa:heptamers}. In 3D one might consider the
distribution of atmospheric or intergalactic dust aggregates. Thus the
polydispersed case represents a very wide class of systems, whose possible
scaling properties and apparent FD are quantities of interest.  Both the BC
and MS functions and FDs will be calculated, first generally, and then for a
number of specific but broadly used radii distributions.

\subsubsection{Minkowski Analysis}
\label{Minkowski}

The Minkowski function and dimension corresponding to
the model of polydispersed spheres introduced above, is calculated next.
\newline{}

\paragraph{The Minkowski Function \newline{}}
Let all lengths be normalized to the total linear extent of the surface $L$.
For a given probability distribution of radii $P(a)$, consider the
determination of the MS function for $M$ randomly placed spheres with a
specific realization of radii out of $P(a)$, ${\bf a} \equiv
(a_1,a_2,...,a_M)$.  The MS function for this realization, $V_{\bf
a}^{(n)}(R)$, is the volume of the union of spheres of radius $R$ centered at
all points in the set under consideration. This amounts to increasing the
intrinsic radius of a sphere $a_i$ to $a_i+R$, and then calculating the volume
of the union. Let the total volume of the embedding space be $V_t$. It is
convenient to work with the normalized volume $\alpha_{\bf a}^{(n)}(R) \equiv
V_{\bf a}^{(n)}(R)/V_t$. The calculation of $\alpha_{\bf a}^{(n)}(R)$ is very
similar to the calculation of the BC function. A point is randomly chosen in
the embedding space, and one calculates the probability, $q_M$, that after
placement of M spheres with radii ${\bf a}$, the chosen point is not included
in the volume of any of the spheres. The probability $q_1$ for this to happen
after random placement of the first sphere is proportional to the volume
remaining after subtracting the volume of this sphere,

\begin{equation}
q_1 = 1-{{\gamma_n(a_1+R)^n} \over V_t},
\label{eq:M:q1}
\end{equation}

\noindent where

\begin{eqnarray}
\gamma_n = \left\{ \begin{array}{ll}
	2                & \mbox{: $n=1$} \\
	\pi     	 & \mbox{: $n=2$} \\
	{4\pi \over 3}    & \mbox{: $n=3$}
	\end{array}
\right.
\label{eq:M:gamma_n}
\end{eqnarray}

\noindent is a geometrical factor associated with the volume of an
$n$-dimensional sphere. The next placements are independent, so that

\begin{equation}
q_M = \prod_{i=1}^{M} \left(1-{{\gamma_n(a_i+R)^n} \over V_t} \right) .
\label{eq:M:qM}
\end{equation}

\noindent Now, $1-q_M$ is the probability of finding the chosen point {\em in}
the set, which on the other hand is equal to the normalized volume of
the set. Thus:

\begin{equation}
\alpha_{\bf a}^{(n)}(R) = 1-\prod_{i=1}^{M}
\left(1-{{\rho\,\gamma_n(a_i+R)^n} \over M} \right) .
\label{eq:M:alpha_a}
\end{equation}

\noindent where $\rho = M/V_t$ is the number density.

Next, average over all possible realizations of radii. Let

\begin{equation}
\langle f(a) \rangle \equiv \int P(a)\,f(a)\,da
\label{eq:M:ave}
\end{equation}

\noindent denote the average of any function of the radius. Then the
expectation value of the normalized volume, when each realization of radii
${\bf a}$ is weighted by its probability $\prod_{i=1}^M P(a_i)\,da_i$, is
given by:

\begin{eqnarray}
\alpha^{(n)}(R) \equiv \langle \alpha_{\bf a}^{(n)}(R) \rangle = 
1-\int \prod_{i=1}^{M} da_i\: P(a_i)\,\left(1-{{\rho\,\gamma_n(a_i+R)^n} \over
M} \right) = \nonumber \\
1-\left[ \int da\: P(a)\,\left(1-{{\rho\,\gamma_n(a+R)^n} \over M} \right)
\right]^M = 1-\left(1-{{\rho\,\gamma_n \langle (a+R)^n \rangle} \over M}
\right)^M .
\label{eq:M:alpha-ave}
\end{eqnarray}

\noindent In the limit of large $M$ one finally obtains:

\begin{equation}
\alpha^{(n)}(R) = 1-e^{-\rho\, \gamma_n \langle (a+R)^n \rangle} ,
\label{eq:M:alpha}
\end{equation}

\noindent and, explicitly for each dimension:

\begin{eqnarray}
\alpha^{(1)}(R) &=& 1-\exp\left[-2\rho(R+\langle a \rangle ) \right] \nonumber
\\
\alpha^{(2)}(R) &=& 1-\exp\left[-\pi \rho(R^2 + 2R\langle a \rangle + \langle
a^2 \rangle ) \right] \nonumber \\
\alpha^{(3)}(R) &=& 1-\exp\left[-{4\pi \over 3} \rho (R^3 + 3R^2\langle a
\rangle + 3R\langle a^2 \rangle + \langle a^3 \rangle ) \right] .
\label{eq:M:alphas}
\end{eqnarray}

\paragraph{The Fractal Dimension \newline{}}
To find the FD according to Eq.(\ref{eq:D_M}), we evaluate the logarithmic
derivative of $\alpha^{(n)}(R)$, which is found to be:

\begin{equation}
{{d\log\left(\alpha^{(n)}(R)/R^n \right)} \over {d\log(1/R)}} =
n-{{\rho \gamma_n \left( \sum_{k=1}^n k\, R^k \langle a^{n-k} \rangle \right)
\exp\left(-\rho\, \gamma_n \langle (a+R)^n \rangle \right)} \over
{1-\exp\left(-\rho\, \gamma_n \langle (a+R)^n \rangle \right)}}
\label{eq:M:dlog(alpha)}
\end{equation}

\noindent In analogy with the discussion in Sec.\ref{penetrable-FD}, one may
expect the cut-offs to be found approximately at

\begin{equation}
R_0 = \langle a \rangle \:\:\:\: {\rm and} \:\:\:\: R_1 = {1 \over 2}
\rho^{-1/n} - \langle a \rangle ,
\label{eq:M:cutoffs}
\end{equation}

\noindent compared to $r_0 = d$ and $r_1 = \rho^{-1/n}-d$ for BC.  The
difference from BC by a factor of $2$ is due to the appearance of 
{\em radii} as
opposed to box-lengths (which are equivalent to diameters). As in
Sec.\ref{range}, we define the FD to be the slope at the estimated
middle-point $R_e = \sqrt{R_0\,R_1}$ of the scaling range:

\begin{equation}
D_{MS}^{(n)} = {{d\log\left(\alpha^{(n)}(R)/R^n \right)} \over {d\log(1/R)}}
\left|_{R=R_e} \right.
\label{eq:M:D_M}
\end{equation}

\noindent which yields:

\begin{eqnarray}
D_{MS}^{(1)} &=& 1-{{\sqrt{\eta_1(1-\eta_1)}} \over {\exp \left(
\eta_1+\sqrt{\eta_1(1-\eta_1)} \right)-1}} = 1-{{\nu_1^{1/2}\,\phi_1^{1/2}}
\over {\exp \left( \eta_1 + \nu_1^{1/2}\,\phi_1^{1/2} \right)-1 }} \nonumber \\
D_{MS}^{(2)} &=& 2-{{\nu_2^{3/4}\,\phi_2^{1/2} + 2\nu_2^{1/2}\,\phi_2}
\over {\exp \left( \eta_2 + \nu_2^{3/4}\,\phi_2^{1/2} + \nu_2^{1/2}\,\phi_2
\right)-1 }} \nonumber \\
D_{MS}^{(3)} &=& 3-{{\left( \lambda^{2/3}\,\mu - (2\mu)^2 \right)^{1/2} +
2\nu_3^{2/3}\, \phi_3 + 3\nu_3^{1/2}\,
\phi_3^{3/2}} \over {\exp \left( \eta_3 + \left( \lambda^{2/3}\,\mu -
(2\mu)^2 \right)^{1/2} + \nu_3^{2/3}\,\phi_3 + \nu_3^{1/2}\, \phi_3^{3/2}
\right)-1}}
\label{eq:M:Ds}
\end{eqnarray}

\noindent where for convenience a number of parameters are defined as follows:

\begin{eqnarray}
\eta_n &=& \rho\,\gamma_n\,\langle a^n \rangle \nonumber \\
\nu_n  &=& \eta_n {{\langle a \rangle^n} \over {\langle a^n \rangle}}
\nonumber \\
\phi_n &=& {1 \over 2} \gamma_n^{1/n}-\nu_n^{1/n} \nonumber \\
\mu  &=& {3 \over 2} \eta_3 {{\langle a^2 \rangle \langle a \rangle} \over
{\langle a^3 \rangle}} \nonumber \\
\lambda &=& \eta_3 { {(\pi \langle a^2 \rangle)^{3/2}} \over
{\gamma_3 \langle a^3 \rangle}} .
\label{eq:M:params}
\end{eqnarray}

\noindent The parameter $\eta_n$ is simply the {\em coverage}
\cite{random-comment5}; the other parameters measure various moments of the
radius distribution.

The expressions for $D_{MS}^{(n)}$ can be simplified somewhat for the case of
a constant radius, whence $\langle a^n \rangle = a^n$, and consequently
$\eta_n = \nu_n$.  Note that the FD contains useful information about the
first $n$ moments of the radius-distribution. In the next section the BC
function of the same model of polydispersed radii is solved for, in order to
compare the MS and BC methods.

\subsubsection{Box-counting Analysis}
\label{poly-BC}

\paragraph{The BC Function \newline{}}
Once again, take all lengths to be normalized to the total linear extent of
the surface $L$.  Repeating the averaging arguments used to arrive at
Eq.(\ref{eq:M:alpha-ave}), combined with the derivations of
Eqs.(\ref{eq:<N>}),(\ref{eq:<N>-2D}),(\ref{eq:<N>-3D}), leads to the following
result for the BC (density-)function in the polydispersed case:

\begin{equation}
N^{(n)}(r) \equiv \langle N_{\bf a}^{(n)}(r) \rangle = {1 \over r^n} \left[ 1-
\left( 1-{\rho \over M} \sum_{k=0}^n \beta_{k n}\,r^k\, \langle a^{n-k}
\rangle \right)^M \right] .
\label{eq:poly:<N>}
\end{equation}

\noindent Here $r$ is the normalized box-length, $\rho = M/V_t$ is the
number-density, and

\begin{eqnarray}
\beta_{k n} = \left\{ \begin{array}{ll}
	\gamma_n         & \mbox{: $k=0$} \\
	1	     	 & \mbox{: $k=n$} \\
	2n	    	 & \mbox{: $k=n-1$} \\
	3\pi	    	 & \mbox{: $k=1$, $n=3$}
	\end{array}
\right.
\label{eq:poly:beta_kn}
\end{eqnarray}

\noindent is introduced for convenience. Taking the limit of large $M$, one
finds:

\begin{equation}
N^{(n)}(r) = {1 \over r^n} \left[ 1- \exp \left( -\rho\, \sum_{k=0}^n \beta_{k
n}\,r^k\, \langle a^{n-k} \rangle \right) \right] .
\label{eq:poly:N-exp}
\end{equation}

\noindent Although at first sight this result may seem to differ substantially
from that for the MS function $\alpha^{(n)}(R)$ (Eq.\ref{eq:M:alpha}), 
the two will be compared in Sec.\ref{BC-vs-M} and shown to be quite
similar.
\newline{}

\paragraph{The Fractal Dimension \newline{}}
Proceeding  with the calculation of the FD, evaluation of the logarithmic
derivative yields:

\begin{equation}
{{d\log\left( N^{(n)}(r) \right)} \over {d\log(1/r)}} =
n-{{\rho \left( \sum_{k=1}^n k\, \beta_{k n}\,r^k \langle a^{n-k} \rangle
\right)} \over {\exp\left(\rho\, \sum_{k=0}^n \beta_{k n}\, r^k \langle
a^{n-k} \rangle \right)-1}} .
\label{eq:poly:dlog(N)}
\end{equation}

\noindent As usual, the FD is defined as the slope at the estimated
middle-point $r_e = \sqrt{r_0\,r_1}$ between the lower and upper cut-offs,
$r_0 = 2\langle a \rangle$ and $r_1 = \rho^{-1/n}-2\langle a \rangle$:

\begin{equation}
D_{BC}^{(n)} = {{d\log\left(N^{(n)}(r)\right)} \over {d\log(1/r)}}
\left|_{r=r_e} \right.
\label{eq:poly:D_BC}
\end{equation}

\noindent with the result:

\begin{eqnarray}
D_{BC}^{(1)} &=& 1-{{\sqrt{\eta_1(1-\eta_1)}} \over {\exp \left(
\eta_1+\sqrt{\eta_1(1-\eta_1)} \right)-1}} = 1-{{\zeta_1^{1/2}\,\psi_1^{1/2}}
\over {\exp \left( \eta_1 + \zeta_1^{1/2}\,\psi_1^{1/2} \right)-1 }} 
\nonumber\\
D_{BC}^{(2)} &=& 2-{{\zeta_2^{3/4}\,\psi_2^{1/2} + \zeta_2^{1/2}\,\psi_2} \over
{\exp \left( \eta_2 + \zeta_2^{3/4}\,\psi_2^{1/2} + {1 \over 2}
\zeta_2^{1/2}\,\psi_2 \right)-1 }} 
\nonumber\\
D_{BC}^{(3)} &=& 3-{{\left( (\lambda')^{2/3}\,\mu - (3\mu)^2
\right)^{1/2} + \zeta_3^{2/3}\,\psi_3 + {1 \over 2} \zeta_3^{1/2}\,
\psi_3^{3/2}} \over {\exp \left( \eta_3 + \left( (\lambda')^{2/3}\,\mu -
(3\mu)^2 \right)^{1/2} + {1 \over 2} \zeta_3^{2/3}\,\psi_3 + {1 \over 6}
\zeta_3^{1/2}\,\psi_3^{3/2} \right)-1 }} ,
\label{eq:poly:Ds}
\end{eqnarray}

\noindent where we defined for convenience:

\begin{eqnarray}
\delta_n &=& \left\{ \begin{array}{ll}
	2                & \mbox{: $n=1$} \\
	8       	 & \mbox{: $n=2$} \\
	48	         & \mbox{: $n=3$}
	\end{array}
\right. \nonumber \\
\zeta_n  &=& {\delta_n \over \gamma_n} \eta_n {{\langle a \rangle^n} \over
{\langle a^n \rangle}} \nonumber \\
\psi_n &=& {1 \over 2} \delta_n^{1/n}-\zeta_n^{1/n} \nonumber \\
\lambda' &=& \eta_3 { {((3\pi)^2 \langle a^2 \rangle)^{3/2}} \over
{\gamma_3 \langle a^3 \rangle}} .
\label{eq:poly:params}
\end{eqnarray}

\noindent Here $\delta_n$ is a geometrical factor associated with the volume of
an $n$-dimensional cube (or box). Note the similarity of
$\zeta_n,\psi_n,\lambda'$ to $\nu_n,\phi_n,\lambda$ respectively of
the previous section.

The BC result for the one-dimensional case is identical to the MS result. This
is due to the fact that in 1D both a box and a sphere reduce to a line
segment. For $n=2,3$ the geometrical factors are different ($\gamma_n$ for the
MS and $\delta_n$ for BC). In Sec. \ref{effect-poly} these issues will be
considered in detail.

\subsubsection{The Distribution Functions}
\label{effect-poly}
In this section the effect of polydispersivity is examined explicitly, by
assuming various functional forms for the radii distributions.

To assess the influence of polydispersivity we considered four types of common
continuous distributions of radii, a well as a simple discrete bi-modal
distribution. These are compared for reference with the case of monodispersed
radii treated for BC in Sec.\ref{basic-model}. The distributions considered
are:

\begin{equation}
\begin{array}{rcll}
P_N(a) &=& {1 \over {\sqrt{2\pi} b}} e^{-(a-\langle a \rangle)^2 \over 2b^2}
& {\rm Normal} \nonumber \\
P_U(a) &=& \left\{ \begin{array}{ll}
	1/b     &  \mbox{: $\langle a \rangle-{1 \over 2} b \leq a \leq
\langle a \rangle+{1 \over 2} b$}\\
	0	&  \mbox{: ${\rm else}$}
	\end{array}
\right.  
& {\rm Uniform} \nonumber \\
P_E(a) &=& {1 \over {\langle a \rangle}} e^{-a/\langle a \rangle} \:\:\:\:\:\:
& {\rm Exponential \cite{random-comment3}} \nonumber \\
P_S(a) &=& {1 \over \Gamma(\langle a \rangle/b)} b^{-\langle a \rangle/b}\,
a^{\langle a \rangle-b \over b}\, e^{-a/b} 
& {\rm Schulz
\cite{random-comment4}} \nonumber \\
P_B(a) &=& \left\{ \begin{array}{ll}
	{\rm Pr}[a=\langle a \rangle-(1-p)b] = p \\
	{\rm Pr}[a=\langle a \rangle+p\,b] = 1-p
	\end{array}
\right. 
& {\rm Bi-modal}
\label{eq:distributions}
\end{array}
\end{equation}

\noindent where $b$ is in all cases a measure of the width of the distribution.
A comparison of the distributions for two combinations of $\langle a \rangle$
and $b$ values is shown in Fig.\ref{fig:distributions}, where they can be seen
to differ significantly. In order to understand the difference between these
distributions in the present context, note that $D_{BC}^{(n)}$ and
$D_{MS}^{(n)}$ are determined by the first $n$ moments of the
distributions. Therefore it is useful to summarize the differences as follows:
let $y_n=\langle a^n \rangle/\langle a \rangle^n$ and $z=b/\langle a
\rangle$. Then:

\begin{equation}
\begin{array}{rcll}
y_2 &=& 1+z^2 \:\:\:\:\: y_3 = 1+3z^2 
& {\rm Normal} 
\nonumber \\
y_2 &=& 1+{z^2 \over 12} \:\:\:\:\: y_3 = 1+{z^2 \over 4} 
& {\rm Uniform}
\nonumber \\
y_2 &=& 2 \:\:\:\:\:\:\:\:\:\:\:\:\:\:\, y_3 = 6 
& {\rm Exponential} 
\nonumber \\
y_2 &=& 1+z \:\:\:\:\:\:\: y_3 = 1+3z+2z^2 
& {\rm Schulz} 
\nonumber \\
y_n &=& p(1-(1-p)z)^n + (1-p)(1+p\,z)^n 
& {\rm Bi-modal}
\label{eq:moments}
\end{array}
\end{equation}

\noindent However, the effect of assuming these different
radii-distributions on the MS and BC functions, and on the respective
FDs is marginal in spite of the differences among them, as shown next.

\subsection{Results}
\subsubsection{Range of Linearity and the Effect of Dimensionality}

In order to meaningfully compare the results in different dimensions, it is
most convenient to fix the average distance $\langle x \rangle \equiv
\rho^{-1/n}$ between sphere centers. Note that this implies different
coverages, since from Eq.(\ref{eq:M:params}): $\eta_n = \gamma_n (\langle a
\rangle/\langle x \rangle)^n$. In particular, since typically $\langle a
\rangle \ll \langle x \rangle$, $\eta_3$ will be much smaller than $\eta_1$
for the same average distance. With this choice, a general estimate for the
range of linearity, independent of the dimension, may be found. Suppose
$\langle x \rangle = 10^k \langle a \rangle$; using the values for the
cut-offs, $R_0=\langle a \rangle$ and 
$ R_1= {1 \over 2} \rho^{-1/n}-\langle a \rangle$, 
one has:\\ $\Delta_n \leq \log R_1 - \log R_0 = \log \left( 
{1 \over 2} \langle x \rangle / \langle a \rangle -1 \right) = \log \left( 
{1 \over 2} 10^k-1 \right) \sim k - 0.3$ for $k>1$.  
As discussed in Sec.\ref{range}, 
the range of linearity is limited both by the distance between cut-offs,
which tends to increase when the coverage decreases and the quality of
the linear regression, measured by ${\cal R}^2$ which tend to improve
as the coverage increases. These trends are independent of the embedding
space dimension and therefore in all dimensions we observe apparent
fractality within a range of 1-2 orders of magnitude.

\subsubsection{Comparison of Box-Counting and Minkowski Sausage Results}
\label{BC-vs-M}
It is comforting to find that, by and large, the BC and MS methods of
resolution analysis yield very similar results. This is shown in
Fig.\ref{fig:BC-vs-M}, where the two methods are compared for the
monodispersed case (quantitatively similar differences result for
poly-dispersivity). When properly normalized (see caption), the methods are
identical in 1D, and differ slightly in 2D and 3D. We shall see below that the
FD values do not differ by more than 0.05 for 2D and 0.1 for 3D either, with
MS giving the consistently lower values
(see Figs. \ref{fig:D-poly.n2.new}, \ref{fig:D-poly.n3.new} below). 
We attribute the small
variance between the two methods to the finite size of the basic building
blocks, and to the differences in the geometrical factors $\gamma_n$ and
$\delta_n$, which determine the details of lowering the resolution of
observation. As the difference is so small, in the next section the entire
discussion is held in terms of BC.

\subsubsection{Effect of Polydispersivity on the BC Function}
\label{effect-poly-funcs}

The BC function is displayed as a function of yardstick size in
Figs.\ref{fig:func-poly.n2.new},\ref{fig:func-poly.n3.new} for the
distributions considered above, for a representative density of spheres,
corresponding to an average intersphere distance $ \langle x \rangle $ of 20
sphere radii (see captions for details). The average radius $\langle a
\rangle$ and width $b$ were set equal for all distributions. The value of $b =
0.9\langle a \rangle$ was chosen so as to reflect a very broad
distribution. Values of $b$ larger than $\langle a
\rangle$ are not permissible since this would lead to negative radii.
An observation of significant importance is that the qualitative (and to a
large extent also the quantitative) nature of the results is unaffected by
changing $b$. The BC function for the 1D case depends only the average radius,
as is clear from Eqs.(\ref{eq:M:alpha-ave}),(\ref{eq:poly:N-exp}), and hence
no difference is observed between distributions with the same $\langle a
\rangle$ in the 1D case. For 2D $\langle a^2 \rangle$ enters, which depends,
in turn, on the specific distribution (Eq.(\ref{eq:distributions})). For 3D
also $\langle a^3 \rangle$ enters, so that a stronger dependence on the
distributions results. However, as seen in
Figs. \ref{fig:func-poly.n2.new},\ref{fig:func-poly.n3.new}, the differences
in BC function for different distributions set in only at small $r$ values,
close to the lower cut-off, and are always bounded.

The two most important questions in the context of apparent fractality relate
to the slopes and the ranges of linearity. The former is dealt with in detail
in the next subsection. As for the range of linearity of the scaling
region, it appears that this is slightly decreased when polydispersivity is
compared to monodispersivity.

\subsubsection{Effect of Polydispersivity on Fractal Dimension}
\label{effect-poly-FDs}

The FD is displayed as a function of coverage in
Figs.\ref{fig:D-poly.n2.new},\ref{fig:D-poly.n3.new} for the distributions of
Eq.(\ref{eq:distributions}). The average radius $\langle a \rangle$ and width
$b$ were set equal for all distributions, with $b=0.9\langle a \rangle$ chosen
again so as to reflect the unfavorable case of a very broad distribution (see
captions for details). Changing among distributions is seen to have only a
minor effect on the FD, for the entire coverage range. A somewhat stronger
effect is observed for 3D than for 2D. For 1D, as seen in
Eqs.(\ref{eq:M:Ds}),(\ref{eq:poly:Ds}), the FD depends only on the average
sphere radius (i.e., rod length), which was taken identical for all
distributions, so that they necessarily all produce the same FD. However, the
second and third moments $\langle a^2 \rangle$, $\langle a^3 \rangle$, which
determine $D^{(2)}$ and $D^{(3)}$ through
Eqs.(\ref{eq:M:Ds}),(\ref{eq:poly:Ds}), do depend on the particular
distribution, and must affect, therefore, the FD. Comparing the moments of
Eq.(\ref{eq:moments}) reveals why the FD is so robust with respect to change
of radius-distribution or its parameters: recalling that $z<1$ in order to
assure positivity of the radii, observe that $y_n$ are typically close to $1$
(the monodispersed case) in all cases, except for the exponential
distribution. (For the Bi-modal distribution it is straightforward to show
that $0<y_n<1$ (recall that $0<p<1$), and is close to zero only for very small
$p$ and $z$.) However, even for the exponential distribution, it is seen in
Figs.\ref{fig:D-poly.n2.new},\ref{fig:D-poly.n3.new} that although the
corresponding FD is indeed somewhat displaced, it is still very close to that
of the other distributions. Thus, it is the combined effect of the relatively
small dependence of the moments on the underlying radius-distribution in the
parameter range of interest, together with the even further suppressed
sensitivity of the FD to these dependences, which is responsible for the
robustness of the FD. The insensitivity of the model to polydispersivity is in
support of our proposition that {\em the random adsorption model is generic:
its features are virtually unaffected by (strong) perturbations in this
commonly encountered way.}

\section{Solution of the Impenetrable Spheres Case}
\label{impenetrable}

\subsection{The Model}
Consider now a different model, which adds correlations on top of the model of
equi-sized, mutually penetrable spheres by imposing {\em im}penetrability on a
system of $n$-dimensional spheres at equilibrium
\cite{random-comment2}. Impenetrability creates a negative correlation in
sphere positions. This model is fully solvable for $n=1$, and approximately
solvable with high accuracy for $n=2,3$. It represents an important class of
processes with correlations, such as models of hard-sphere liquids,
energy-level repulsion in quantum systems which are classically chaotic
\cite{berry}, Langmuir-type adsorption, etc. As demonstrated below, the
correlation due to impenetrability merely {\em modifies} the apparent fractal
character already induced by the random nature of the problem. We limit
ourselves to a BC analysis of this problem.

\subsection{Derivation for 1D Case}

Consider first the 1D case of impenetrable rods: BC function is calculated by
employing a result derived using thermodynamic arguments by Helfand et
al. \cite{Helfand}, and using a statistical argument by Torquato et
al. \cite{Torquato:NN}. They show that the probability of finding a cavity of
length $l$ containing no rod centers, in a system of a large number of
impenetrable rods of length $d$ each, is:

\begin{equation}
q^{(1)} = (1-\eta_1) \exp \left[{-\eta_1 \over (1-\eta_1)} \left({l \over d}-1
\right)\right]
\label{eq:qhard}
\end{equation}

\noindent for $l>d$. In performing BC, for a box of length $r$ to remain empty,
no rod-center may fall within $d/2$ from either side of the box, so that
$l=r+d$. $q^{(1)}$ may be rewritten, in the limit of a large number of rods ($M
\rightarrow \infty$ at fixed coverage), as:

\begin{equation}
q^{(1)} = (1-\eta_1) \left(1- \left({l \over d} -1 \right) {\eta_1 \over
{1-\eta_1}} {1 \over M} \right)^{M} = (1-\eta_1) \left(1- {r \over {1-\eta_1}}
\right)^{M} .
\label{eq:qhard-new}
\end{equation}

\noindent Having $p=1-q$ and using Eq.(\ref{eq:N}) the expected number of
intersected boxes is:

\begin{equation}
\langle N(r)\rangle = {1 \over r} \left(1-(1-\eta_1) \left( 1- {r \over
{1-\eta_1}} \right)^{M} \right) .
\label{eq:Nhard}
\end{equation}

\subsubsection{Fractal Dimension}
As in the penetrable rods case, one can now use Eq.(\ref{eq:D_BC})
(with the slope calculated at $r=r_e$) to calculate a lower bound for the
FD. We obtain:

\begin{equation}
D = 1-\eta_1 \sqrt{{1 \over \eta_1}-1} {{ \left( 1- \sqrt{ {\eta_1 \over
{1-\eta_1}}} {1 \over M} \right)^{M-1}} \over {1-(1-\eta_1) \left( 1- \sqrt{
{\eta_1 \over {1-\eta_1}}} {1 \over M} \right)^{M}}}.
\label{eq:Dhard-full}
\end{equation}

\noindent This can again be simplified for large $M$:

\begin{equation}
D = 1- { {\eta_1 \sqrt{{1 \over \eta_1} -1}}
\over {\exp \left(\sqrt{ {\eta_1 \over {1-\eta_1}}} \right) - (1-\eta_1) } } .
\label{eq:Dhard}
\end{equation}

\subsection{Derivation for 2D and 3D Cases}
So far all the results were based on exact calculations.  We now consider an
approximate solution for 2D and 3D impenetrable spheres. Full analytical
solutions are at present impossible: exact results for the probability of
finding a cavity containing no disk or ball centers after their placement, as
employed in Eq.(\ref{eq:qhard}) above, are not available because the
$n$-particle probability densities are not exactly known. Nevertheless,
Refs.\cite{Torquato:NN,Torquato:hard-spheres} provide some accurate
approximations for the probability of finding a two- or three-dimensional
spherical cavity of radius $l$ in an equilibrium system of hard spheres of
radius $a$.

\subsubsection{BC Function}
In the context of BC, one in fact requires the probability of finding a cavity
with the shape of the convolution of a box and sphere (see
Sec.\ref{basic-model} and Fig.\ref{fig:rods}(d)). This will be undertaken in a
future study; at present we will settle for an approximation of the cavity by
spheres. The important quantity to conserve in this approximation, is the
cavity volume, since this is what actually enters the probabilistic argument
at the root of the calculation of the BC function. If one simply takes the
spherical cavity radius as the geometric mean of the sphere radius plus half
box length, and sphere radius plus half box diagonal,

\begin{equation}
l_n(r) = \left[ \left({r \over 2} + a \right)\left({1 \over 2} \sqrt{n} + a
\right) \right]^{1/2} 
\label{eq:radii}
\end{equation}

\noindent the real cavity volume (see Eqs.(\ref{eq:q1-2D},\ref{eq:q1-3D})) is
overestimated by no more than 10\% in 2D, and (except for a sharp maximum of
80\% for $0.2<r/a<5$) by no more than 20\% in 3D. This will suffice for the
present purpose of an approximate treatment of the 2D and 3D impenetrable
spheres problem. Now, the results of Ref.\cite{Torquato:NN} for the
empty-cavity probability in $n$ dimensions can be expressed conveniently as
follows:

\begin{equation}
q^{(n)}(r) = (1-\eta_n) \exp \left( {-\eta_n \over (1-\eta_n)^n} f_n(r)
\right) ,
\label{eq:hard-qn}
\end{equation}

\noindent where the subscript $n$ on the coverage $\eta$ serves to remind that
the same coverage defines different combinations of sphere sizes and densities
for different dimensions (see Eq.(\ref{eq:M:params}) in the monodispersed
case). The functions $f_n$ depend on the particular approximation used, but
have the general form

\begin{equation}
f_n(r) = \sum_{j=0}^n \alpha_j^{(n)}(\eta_n) \, x_n^j; \:\:\:\:\:\:\: \: x_n =
{l_n(r) \over 2a} .
\label{eq:hard-fn}
\end{equation}

\noindent The $\alpha_j^{(n)}$ are, for $n=2$ (impenetrable disks), using the
scaled-particle theory of Reiss, Frisch and Lebowitz \cite{Reiss}:

\begin{equation}
\alpha_0^{(2)} = 2\eta-1 \:\: ;\: \alpha_1^{(2)} = -4\eta \:\: ;\:
\alpha_2^{(2)} = 4 ,
\label{eq:hard-alpha-2D}
\end{equation}

\noindent whereas for $n=3$, using the Carnahan-Starling result
\cite{Carnahan}, Ref.\cite{Torquato:NN} finds:

\begin{equation}
\alpha_0^{(3)} = -{1 \over 2} (9\eta^2 - 7\eta +2) \:\: ;\: \alpha_1^{(3)} =
12\eta^2
\:\: ;\: \alpha_2^{(3)} = -6\eta(3+\eta) \:\: ;\: \alpha_3^{(3)} = 8(1+\eta) .
\label{eq:hard-alpha-3D}
\end{equation}

\noindent Next, the BC functions are given, as usual, by:

\begin{equation}
\langle N^{(n)}(r) \rangle = {1 \over r^n} \left( 1-q^{(n)} \right) ,
\label{eq:N-hard}
\end{equation}

\noindent The apparent FDs predicted by these results are discussed next.

\subsubsection{Fractal Dimension}
For the 2D and 3D cases, the FD is obtained by evaluating the logarithmic
derivative at $r_e^{(n)} = \left[ 2a \left( \rho^{-1/n} - 2a \right)
\right]^{1/2}$. Using Eq.(\ref{eq:M:params}), the estimated middle-point of
the scaling range can be rewritten as:

\begin{equation}
{r_e^{(n)} \over a} = 2 h_n(\eta) \:\: ; \:\:\:\:\: h_n(\eta) = \left[
{1 \over 2}
\left( {\gamma_n \over \eta_n} \right)^{1/n} - 1 \right]^{1/2} .
\label{eq:re/a}
\end{equation}

\noindent The hard-sphere radius $a$ cancels out and the FD for $n=2,3$ is
found to be given by:

\begin{eqnarray}
D^{(n)} = {d(\log[N^{(n)}(r)]) \over d(\log[1/r])}\left|_{r=r_e}\right. =
n-{1 \over 2} {\eta_n \over (1-\eta_n)^{n-1}} h_n(\eta_n)
[\sqrt{n}(2h_n(\eta_n)+1)+1] \times \nonumber \\
{ { \sum_{j=1}^n 2^{-j} \alpha^{(n)}_j(\eta_n) \, j \, [ (1+h_n(\eta_n))
(1+\sqrt{n}h_n(\eta_n)) ]^{j/2-1}} \over {\exp \left[ {\eta_n \over
(1-\eta_n)^n} \sum_{j=0}^n 2^{-j} \alpha^{(n)}_j(\eta_n) [(1+h_n(\eta_n))
(1+\sqrt{n} h_n(\eta_n))]^{j/2} \right] - (1-\eta_n)} } .
\label{eq:D_n-hard}
\end{eqnarray}

\noindent It should be noted that $D^{(n)}$ are functions of $\eta_n$ alone,
which indeed, in contrast to the penetrable spheres case, is exactly the
volume fraction of space occupied by the impenetrable spheres.

\subsection{Results}

\subsubsection{Effect of Impenetrability on the BC Function and on the Range of
Linearity}
\label{effect-imp-funcs}

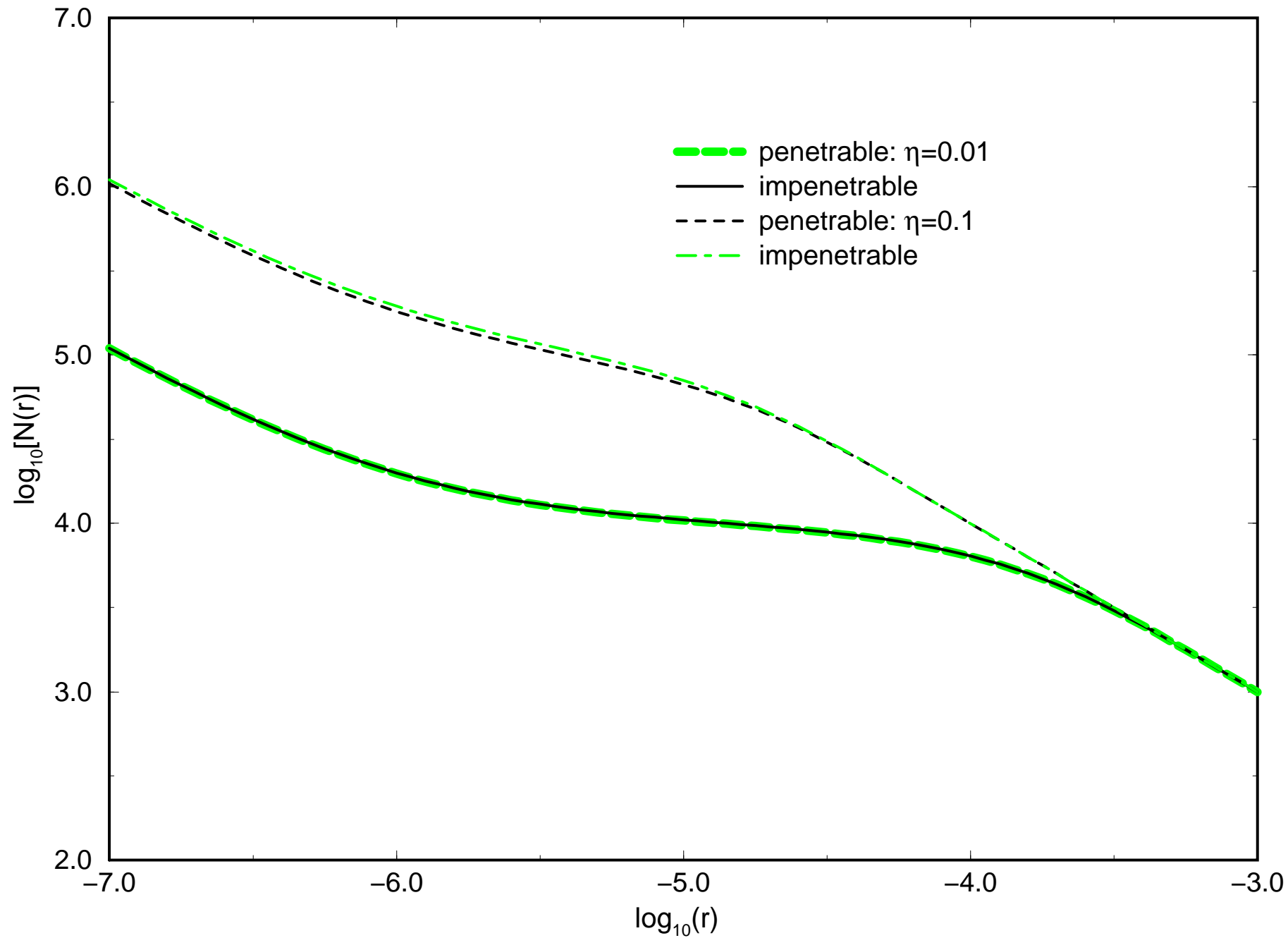
\begin{figure}
\caption{Comparison of box-counting predictions in penetrable and impenetrable
rods cases. The results for penetrable (Eq.(\protect\ref{eq:<N>})) and
impenetrable rods (Eq.(\protect\ref{eq:Nhard})) virtually coincide for $\eta
\leq 10^\protect{-2\protect}$. For $\eta=0.1$ a barely noticeable difference
develops. $d=10^\protect{-6\protect}$ in both cases.}
\label{fig:N.soft-vs-hard}
\end{figure}

Fig.\ref{fig:N.soft-vs-hard} shows the plot of Eq.(\ref{eq:<N>}) for the
penetrable rods case , together with the expression for $\langle N(r)\rangle$
in the 1D impenetrable rods case, Eq.\ref{eq:Nhard}. The behavior is
qualitatively similar in both cases, and virtually indistinguishable for low
coverages. Fig.\ref{fig:N.soft-vs-hard} thus {\em demonstrates the primary
role of pure randomness in the appearance of fractality, even in the presence
of correlations}, at least at coverages below 10\%.

The impenetrability case results for 2D and 3D are shown in
Figs.\ref{fig:D-poly.n2.new} and \ref{fig:D-poly.n3.new} respectively, along
with those for penetrable spheres. As in the 1D case, the effect of
impenetrability is virtually unnoticeable in terms of the FD. Once more, it
appears that {\em the scaling content of this highly non-trivial system is
already contained to a very large extent in its penetrable counterpart}.

\subsubsection{Effect of Impenetrability on Fractal Dimension}
\label{effect-imp-FDs}

The result for 1D is shown in Fig.\ref{fig:D}. As expected, the FD rises to
$1$ faster than in the penetrable rods case: when no overlap is allowed, the
line is filled up at lower coverages. The important observation is the large
range of coverages for which the penetrable and impenetrable rods FDs nearly
overlap. For all practical
purposes, therefore, the FD is very close in these two cases, which differ
significantly in the extent of correlations present in the respective systems:
the fractal content of the impenetrable rods geometry is already contained to
a large extent in the associated random penetrable rods case.

\section{Conclusions}
\label{conclusions}

We have shown in this paper that randomness in its most elementary forms,
generates apparent fractal structures over 1-2 decades, between a lower
cut-off -- the elementary building block -- and an upper cut-off which is
approximately the average distance between building blocks. We adopted an
empirical-like approach to the calculation of fractal dimension: we
deliberately considered sets bound within finite cut-offs, which display
scaling behavior over {\em a physically relevant range}. We believe this
approach to be both useful and necessary, if a direct contact between theory
and experiment is to be achieved. This led us to consider several simple, but
widely applicable models of random phenomena, with and without correlations,
obtaining as one of the main results, an analytical solution for the apparent
fractal dimension of models of randomness. The models studied are convenient
starting points for other, more elaborate ones of random phenomena. It is
argued that the models we have studied are, in fact prototypical of a large
class of spatial, temporal and even spectral random phenomena. The methods
introduced here should be useful in the study of further cases of random
phenomena, with other types of disorder and correlations. To conclude, we
would like to point out that apparent fractal behavior is expected to occur
for a more general class of distributions. Generally, the $\log N(r)$
vs. $\log(1/r)$ plot includes lines of slope $D=n$ ($n=1,2,3$) beyond the
upper and lower cut-offs which are connected by an interval of slope $D<n$,
which depends on the specific distribution.\\\\

\noindent {\bf Acknowledgements}\\
We would like to thank P. Bak, R.B. Gerber, C. Jayaprakash, D. Mukamel,
J. Sethna and G. Shinar for very helpful discussions.  D.H. and D.A. are
members of the Fritz Haber Research Center for Molecular Dynamics. D.A. is
also a member of the Farkas Center for Light Energy Conversion.

\noindent
\begin{figure}
\caption{Typical realizations of random placements of rods of size $d$ (gray
with center dot) in the penetrable (a) and impenetrable (b) cases. Also shown
is a division into ``boxes'' (vertical lines). The figure also represents the
zeroset of a temporal series of signals of width $d$, or a spectrum of
randomly positioned energy levels with uncertainty-width $d$. (c) Illustration
of counting procedure used in our probabilistic arguments: For a box of length
$r$ to remain unintersected, no rod center may approach its ends closer than a
distance of $d/2$. Hence a total length of $r+d$ must remain free. For
Sec.\protect\ref{2D3D-extension}: (d) Excluded area in the case of placement
of disks in 2D. The excluded area consists of the box of side length $r$, four
quarter circles of diameter $d$, and four rectangles $r\cdot d$. This is just
the convolution of the box and a circle of diameter $d$. (e) Counting
procedure in Sec.\protect\ref{correlations}: Distance between the ``test
points'' $x$ is larger than $d$: for both points $x_0$ and $x_0+x$ to be
empty, a total length of $2d$ must be excluded. (f) If $x<d$, a rod falling in
between the two points may overlap both, so a length of only $x+d$ is
excluded. }
\label{fig:rods}
\end{figure}

\noindent
\begin{figure}
\caption{Comparison of simulation results (circles) to the theoretical
prediction of Eq.(\protect\ref{eq:<N>}) (solid line) for the number of
intersected boxes as a function of their size in the 1D penetrable rods
case. The coverage is $\eta=0.1$ and the rod length is
$d/L=10^\protect{-5\protect}$. The cut-offs are manifested as the two knees in
the graph. The lower bound $r_0$ is seen to be indeed located at $r=d$. The
upper bound $r_1$ is at $r=d/\eta-d$, also conforming with the prediction in
the text. The agreement between theory and simulations is excellent over the
entire range. Inset: Same with $\eta=0.01$ and
$d/L=10^\protect{-6\protect}$. Note the increase in the range of linearity.}
\label{fig:Nresults}
\end{figure}

\noindent
\begin{figure}
\caption{The simulation results (circles) for the number of intersected boxes
$N(r)$ vs. $r$ in the experimentally relevant range are shown along with a
linear regression fit for coverage $\eta = 0.1$ (obtained for $d=10^{-5}$,
$N_d = 10^4$). This is the experimentally relevant range which is typically
used to obtain the FD.}
\label{fig:Nresults.reg}
\end{figure}

\noindent
\begin{figure}
\caption{The range of linearity, $\Delta$, of Eq.(\protect\ref{eq:D_BC}), in a
linear regression analysis. The range of linearity decreases as higher quality
regression is required (see text). The results presented are valid in all
dimensions, but it should be remembered that the same coverage corresponds to
different inter-particle distances in different dimensions.}
\label{fig:range}
\end{figure}

\noindent
\begin{figure}
\caption{Apparent fractality (FD) as computed by linear regression with a
relatively high (0.995) coefficient of determination, in the case of rods. The
prediction of the analytical equations, Eqs.(\protect\ref{eq:Dresult}) and
(\protect\ref{eq:Dhard}) serve as accurate lower bounds. The prediction for
the FD of the penetrable and impenetrable rods cases is seen to differ only
marginally, indicating that the dominant contribution to the FD comes from the
penetrable (i.e., totally random) rods case. The lowest coverage shown
corresponds to the lowest molecular densities observed in nature: $10\AA/{\rm
cm}^3$ in intergalactic space.}
\label{fig:D}
\end{figure}

\noindent
\begin{figure}
\caption{Probability distributions $P(a)$ used for polydispersivity in radii of
penetrable spheres model: Normal, uniform, exponential, Schulz, bi-modal. All
distributions have the same average $\langle a \rangle$ and width $b$, defined
in the text. Shown here is the wide case, $\langle a \rangle=1,\: b=0.9$.}
\label{fig:distributions}
\end{figure}

\noindent
\begin{figure}
\caption{Comparison of box-counting ($\log[N(r)]$ vs $\log(r)$) and Minkowski
($\log[\alpha(R)]$ vs $\log(R)$) functions for monodispersed, mutually
penetrable spheres in $n=1,2,3$ dimensions. The
linear density is fixed at $\rho_1 = 0.05/a$, where $a=1$ is the
sphere-radius. A meaningful comparison is achieved by plotting the Minkowski
function normalized to the sphere volume ($\gamma_n R^n$), with $R=r/2$. In 1D
BC and MS are identical. Differences do develop, albeit for ruler sizes ($r$,
$R$) beyond the cut-offs, for $n=2$ and $n=3$, which depend on $\langle a^2
\rangle$ and $\langle a^2 \rangle,\langle a^3 \rangle$ respectively, and have
different geometrical factors due to the use of boxes and M-spheres. Note that
$N(r)$ and $\alpha(R)$ are normalized to the total number of boxes and the
total volume respectively.}
\label{fig:BC-vs-M}
\end{figure}

\noindent
\begin{figure}
\caption{BC functions for poly-dispersed, mutually penetrable spheres in $n=2$
dimensions. As in Fig.\protect\ref{fig:BC-vs-M}, the linear density is $\rho_1
= 0.05/\langle a \rangle$ (note that $\rho$ has dimensions of inverse
area). The distribution parameters are $\langle a \rangle=1,\: b=0.9$. As seen
clearly in the inset, the exponential distribution lies highest, followed by
Schulz, normal, bi-modal, uniform, and finally the monodispersed case (no
distribution). Since the lower cut-off is expected at $\sim \log 1=0$, it
appears that the effect of polydispersivity is to decrease the range of
linearity somewhat.}
\label{fig:func-poly.n2.new}
\end{figure}
 
\noindent
\begin{figure}
\caption{Same as Fig.\protect\ref{fig:func-poly.n2.new}, but in $n=3$
dimensions ($\rho$ has dimensions of inverse volume). The asymmetry between
the lower and upper cut-offs is apparent.}
\label{fig:func-poly.n3.new}
\end{figure}

\noindent
\begin{figure}
\caption{Fractal dimension, obtained as slope of the BC and Minkowski functions
at the estimated middle points $r_e$, $R_e$, for polydispersed, mutually
penetrable spheres in 2D, as a function of the coverage,
$\log_\protect{10\protect}\eta_2$. The minimal coverage is as in
Fig.\protect\ref{fig:D}. Displayed here are the broad-distribution results:
$\langle a \rangle = 1$, $b=0.9\langle a \rangle$. Small differences are
observed among the distributions. The order is opposite to that in
Fig.\protect\ref{fig:BC-vs-M}, with the Minkowski dimension consistently
somewhat smaller. Also shown is the FD for impenetrable disks, which as
expected, is slightly larger. The lowest coverage shown corresponds to the
lowest molecular densities observed in nature: $10\AA/\protect{\rm
cm\protect}^3$ in intergalactic space.}
\label{fig:D-poly.n2.new}
\end{figure}

\noindent
\begin{figure}
\caption{\noindent Fractal dimension, obtained as slope of the 
BC and Minkowski function in 3D. The details are the same as in 
Fig.\protect\ref{fig:D-poly.n2.new} and  
the results are qualitatively the same as in 2D.}
\label{fig:D-poly.n3.new}
\end{figure}

\noindent


\begin{thebibliography}{10}
\bibitem{D:histogram}
{In Ref.\cite{me:D}, we include a histogram, based on all the experimental FD
  measurements presented in Ref. \cite{Feder}, and showing the number of plots
  as a function of the number of decades of the linear range. The conclusion is
  that by and large, most experimental measurements of fractal dimensions are
  based on data that extends between one and two decades.}
\bibitem{random-comment4}
{The Schulz distribution is a continuous version of the discrete Gamma
  distribution. Its standard formula is $P_S(a) = {1 \over \Gamma(m+1)} \left(
  {m+1 \over {\langle a \rangle}} \right)^{m+1} a^m\, e^{-(m+1)a/{\langle a
  \rangle}}$ (G. V. Schulz, Z. Phys. Chem. B {\bf 43}, 25 (1939)), where $m>-1$
  determines the peakedness of the distribution. A general expression for the
  $n$th moment is: $\langle a^n \rangle = b^n \prod_{i=1}^n \left({a-b \over
  b}+i \right)$. The parameter $b$ we introduced is a measure of the width of
  the distribution: $b=\langle a \rangle-a^*$, where $a^*=\langle a \rangle {m
  \over m+1}$ is the position of the maximum of $P_S(a)$.}
\end{thebibliography}
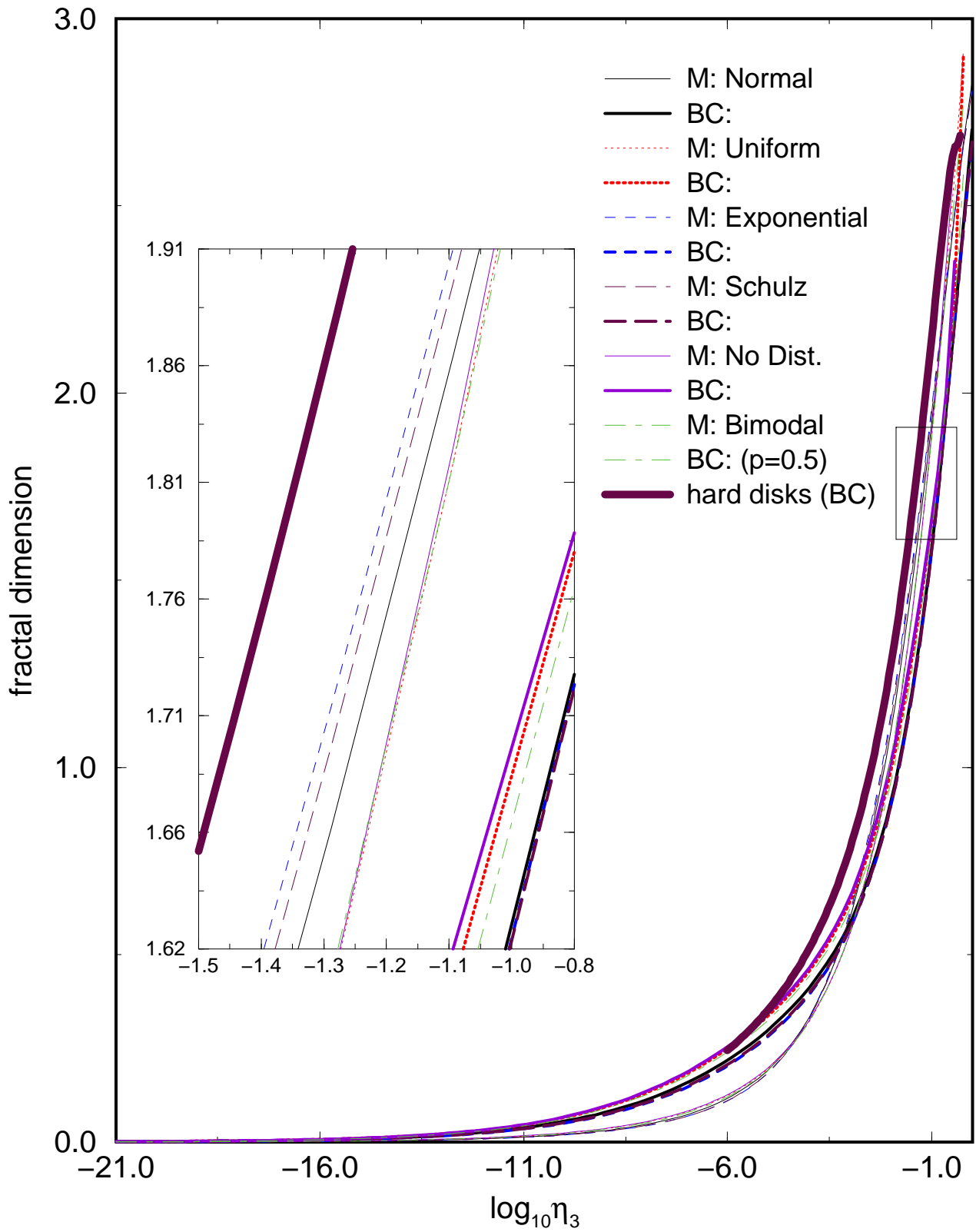
\end{document}